\documentclass[prl,aps,amsfonts,amsmath,amssymb,nofootinbib,twocolumn,linenumbers,superscriptaddress]{revtex4}

\usepackage[switch]{lineno}

\usepackage{graphicx}
\usepackage{bm}
\usepackage{hyperref}

\usepackage{natbib}
\usepackage{float}
\usepackage{xcolor}
\usepackage{gensymb}

\restylefloat{table}

\begin{document}

\title{Dynamical crossover between stretched- and compressed-exponential relaxation in a photoexcited crystal}

\author{Tyler Carbin}
\email{tcarbin@g.ucla.edu}
\affiliation{Department of Physics and Astronomy, University of California Los Angeles, Los Angeles, CA 90095, USA}

\author{Joel Herman}
\affiliation{Department of Physics and Astronomy, University of California Los Angeles, Los Angeles, CA 90095, USA}

\author{Xinshu Zhang}
\affiliation{Department of Physics and Astronomy, University of California Los Angeles, Los Angeles, CA 90095, USA}

\author{Adrian B. Culver}
\affiliation{Department of Physics and Astronomy, University of California Los Angeles, Los Angeles, CA 90095, USA}
\affiliation{Mani L. Bhaumik Institute for Theoretical Physics, Department of Physics and Astronomy, University of California Los Angeles, Los Angeles, CA 90095}

\author{Yu Zhang}
\affiliation{Department of Physics, University of Colorado at Boulder, Boulder, Colorado 80309, USA}

\author{Hengdi Zhao}
\affiliation{Department of Physics, University of Colorado at Boulder, Boulder, Colorado 80309, USA}

\author{Rahul Roy}
\affiliation{Department of Physics and Astronomy, University of California Los Angeles, Los Angeles, CA 90095, USA}
\affiliation{Mani L. Bhaumik Institute for Theoretical Physics, Department of Physics and Astronomy, University of California Los Angeles, Los Angeles, CA 90095}

\author{Gang Cao}
\affiliation{Department of Physics, University of Colorado at Boulder, Boulder, Colorado 80309, USA}

\author{Anshul Kogar}
\email{anshulkogar@physics.ucla.edu}
\affiliation{Department of Physics and Astronomy, University of California Los Angeles, Los Angeles, CA 90095, USA}

\begin{abstract}
    Anomalous relaxation is one of the hallmarks of disordered systems. 
Following perturbation by an external source, many glassy, jammed and amorphous systems relax as a stretched or compressed exponential as a function of time. 
However, despite their ubiquity, the origins of and the connection between these phenomenological relaxation functions remains to be understood.
Here, we observe a tunable crossover from stretched- to compressed-exponential relaxation by photoexciting single crystal Ca$_3$Ru$_2$O$_7$ across a structural phase transition.
%
We present a simple lattice model that shows how spatial inhomogeneity and local, strain-mediated interactions cooperate to produce the dynamical crossover. 
Our work reveals anomalous relaxation dynamics in an idealized single crystal material and establishes photoexcited solids as promising platforms for probing the mechanisms underlying anomalous relaxation.
\end{abstract}

\date{\today}

\maketitle

The stretched/compressed exponential (also known as the Kohlrausch–Williams–Watts function) is a phenomenological function that is used to fit relaxation data across a wide range of disciplines. 
Because these dynamics are nearly universally observed in glasses and supercooled liquids~\cite{Ediger1996,Angell00}, they are commonly referred to as ``glassy".
But, stretched/compressed-exponential relaxation is regularly reported in disordered systems ranging from soft materials such as colloidal suspensions, emulsions, and gels~\cite{Evans_Cates_2000}, to astrophysical processes~\cite{DOBROVOLSKIS2007}, protein folding~\cite{METZLER1998}, and internet traffic patterns~\cite{internet}. 
Due to the ubiquity of this relaxation function, understanding its physical origins is of great fundamental significance. 
A key question is how stretched- and compressed-exponential relaxation are related. 
Although the two take the same mathematical form (see below), 
they are thought to indicate vastly different physics.
Stretched exponentials are often attributed to diffusion traps, heterogeneous relaxation or hierarchically-constrained microscopic dynamics~\cite{Phillips96,Palmer84,Guo09,Johnston2006}, whereas compressed exponentials are associated with ballistic or avalanche-like dynamics~\cite{Cipelletti03,Ferrero14,Evenson15,Trachenko21,Song22}. 
Nonetheless, a number of systems have been found to exhibit both types of anomalous relaxation as a function of a thermodynamic variable, such as temperature~\cite{Ruta12,Evenson15,Luo17,Caronna08,Guo09}. These observations suggest that, in certain cases, the two behaviors may represent different aspects of a single underlying phenomenon.
A promising approach for studying glassy dynamical systems is therefore to investigate the relationship between stretched- and compressed-exponentials by identifying systems that can be tuned between the two regimes.

Solids excited by ultrafast light pulses are one class of systems that offers promise for investigating glassy relaxation.
Generally, to study relaxation dynamics, a perturbation must be applied that is much faster than the intrinsic response time of the system.
In the case of glass, this perturbation is typically a thermal quench, whereas in ultrafast experiments it takes the form of a femtosecond laser pulse, or an ``optical quench".
Here, we investigate the photoinduced phase transition (PIPT) in Ca$_3$Ru$_2$O$_7$ (CRO)~\cite{Cao97}, grown by the floating-zone method~\cite{Supp}.
We choose this compound because it exhibits 
mesoscopic phase coexistence following photoexcitation at low temperatures~\cite{bootstrap}, opening the possibility for anomalous relaxation despite its crystalline structure. 
As CRO is cooled, it first undergoes a continuous phase transition at $T_{N}$=56~K to an antiferromagnetic state in which the Ru spins are aligned ferromagnetically along the $\pm a$ axis within each bilayer and antiferromagnetically between bilayers~\cite{Bao2008}. 
Of primary interest here is the discontinuous metal-insulator transition at $T_{MI}$=48~K.
This electronic transition is accompanied by a rotation of the spins from the $\pm a$ axis to the $\pm b$ axis~\cite{Bao2008,Bohnenbuck08} and a structural transition that preserves the crystallographic symmetry. Through the transition, the $c$ axis contracts by $\sim$0.1\% and the $a$ and $b$ axis lattice parameters expand by $\sim$0.07\%~\cite{Yoshida05}. 

To investigate the relaxation dynamics following photoexcitation, we perform time-resolved second harmonic generation (SHG) in a reflection geometry, as depicted in Fig.~\ref{Fig1}(a). 
The power of this technique in studying the PIPT in CRO is twofold.
First, the SHG probe is directly sensitive primarily to the change in the crystal structure across the metal-insulator transition, which significantly simplifies the interpretation of time-resolved data~\cite{bootstrap}.
Furthermore, rotational anisotropy (RA) SHG provides symmetry information, which enables us to rule out the effects that arise from 90$^\circ$ polar domains.
For both the pump and probe beams, we used a laser centered at 1030~nm (1.20~eV) with 180~fs pulses and a 5~kHz repetition rate. The probe was shone normal to the $ab$-plane with a 35~$\mu$m spot size (FWHM) and a 7.3~mJ/cm$^2$ fluence. 
The pump was shone obliquely on the sample at 15$^\circ$ to the surface normal with a 210~$\mu$m FWHM.

\begin{figure}
    \centering
    \includegraphics[scale=0.9]{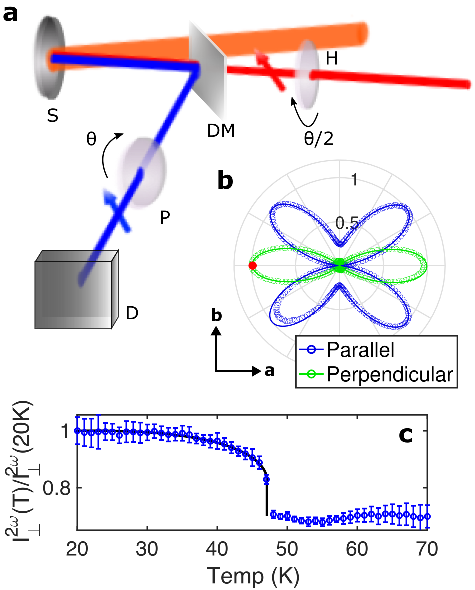}
    \caption{Equilibrium SHG characterization of Ca$_3$Ru$_2$O$_7$. \textbf{(a)} Illustration of time-resolved SHG experimental setup. SHG light is generated and reflected at the CRO sample (S) and is collected at (D), which represents a photomultiplier tube and optical filters to isolate the frequency doubled light. The half-wave plate (H) and polarizer (P) are rotated simultaneously to select the polarization of the incident and outgoing light, respectively. The dichroic mirror assembly (DM) separates the reflected fundamental and frequency-doubled beams. The pump beam (orange) is incident at 15$^{\circ}$ to the surface normal. \textbf{(b)} Rotational anisotropy SHG measured at room temperature. The blue(green) circles are the SHG intensity $I^{2\omega}$ measured as the light polarization is rotated with parallel(perpendicular) incident and outgoing polarizations. The crystallographic axes $a$ and $b$ are indicated by the black arrows. The solid lines are a simultaneous fit to both polarization channels~\cite{Supp}. The data are normalized to a maximum in the perpendicular channel. \textbf{(c)} SHG intensity $I^{2\omega}_{\perp}$ as a function of temperature across the metal-insulator transition $T_{MI}$ normalized to its value at 20~K. The polarization is fixed to the maximum in the perpendicular channel indicated by the red dot in \textbf{b}, such that $I^{2\omega}_{\perp}\propto|\chi_{baa}|^2$. The black line is a guide to the eye.}
    \label{Fig1}
\end{figure}

Equilibrium rotational anisotropy SHG patterns measured at room temperature are shown in Fig.~\ref{Fig1}(b).
Data was collected in two configurations -- one with parallel and one with perpendicular incident and outgoing light polarization. 
We fit both channels simultaneously with the leading order electric dipole SHG contribution, depicted by the solid lines~\cite{Supp}. 
Upon cooling below $T_{MI}$, the symmetry of the rotational anisotropy SHG is unchanged~\cite{bootstrap}, but there is a discontinuous increase in the SHG intensity below $T_{MI}$, as shown in Fig.~\ref{Fig1}(c).
%
%
We observe no thermal hysteresis, consistent with previous work~\cite{Cao97,Sokolov2019, bootstrap}. 

We now study the PIPT instigated by femtosecond light pulses. 
When the system is photoexcited below $T_{MI}$, the SHG decreases in intensity with no change in the symmetry of the rotational anisotropy, as shown in Fig.~\ref{Fig2}(a). 
In Fig.~\ref{Fig2}(b) we plot the relative change in the SHG, $I^{2\omega}(t)/I^{2\omega}(t<0)$, following photoexcitation with varying fluence. 
Remarkably, despite the drastic variation in the timescale and curvature, we obtain a good fit to each time trace with the phenomenological stretched/compressed-exponential function: 
\begin{equation}
\label{Eqn1}
    u(t) = 1+\theta(t)I_0\mathrm{exp}[-(t/\tau)^\beta]
\end{equation}
where $\theta(t)$ is the Heaviside step function.
To corroborate the agreement between the data and this functional form, we rearrange this equation to determine a scaling of the data that collapses all of the time traces to a single line, shown in Fig.~\ref{Fig2}(d).  
The substantial overlap of these curves indicates the degree to which they adhere to this single functional form. 
See the Supplementary Materials for details of the fitting procedure~\cite{Supp}.

\begin{figure*}
    \centering
    \includegraphics[scale=0.37]{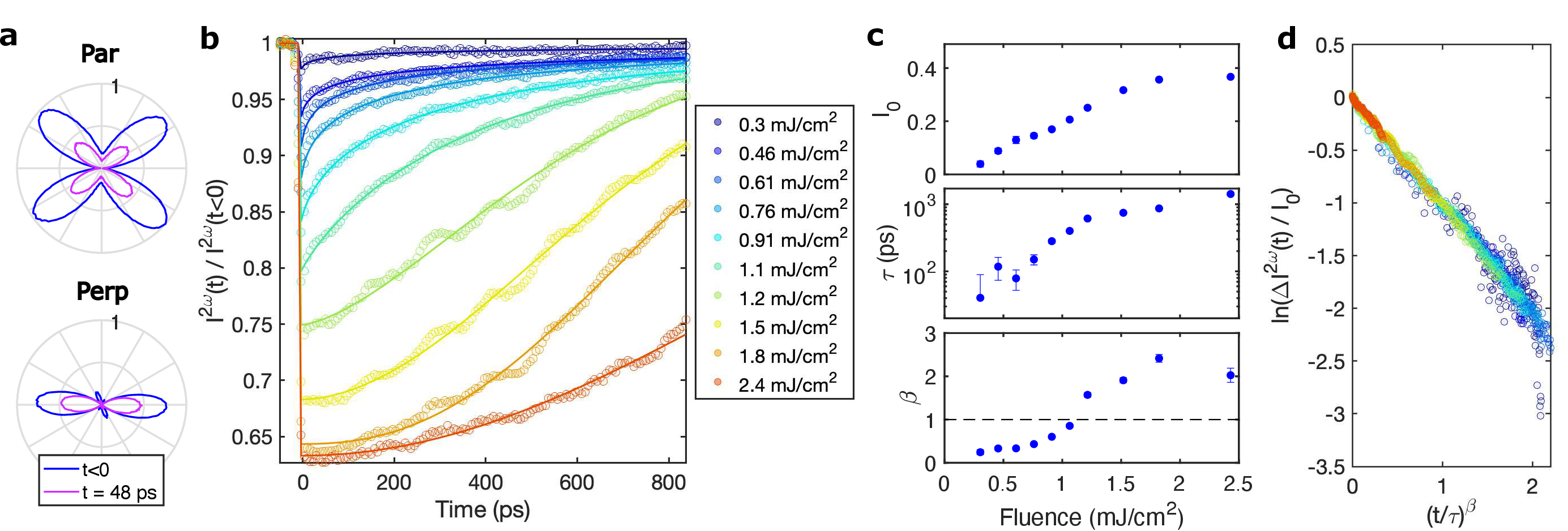}
    \caption{Crossover from stretched- to compressed-exponential relaxation in the PIPT. \textbf{(a)} The rotational anisotropy SHG before (blue) and after (purple) photoexcitation at 4~mJ/cm$^2$ and $T=$4~K. \textbf{(b)} Time evolution of the normalized SHG intensity $I^{2\omega}$ following photo-excitation for varying pump fluence at a nominal temperature of 4~K (laser heating raises the temperature). Solid lines are fits to equation~(\ref{Eqn1}). \textbf{(c)} Best-fit values for each of the fit parameters in equation~(\ref{Eqn1}) as a function of photo-excitation fluence. Error bars are 95\% confidence intervals from the fitting procedure. 
    \textbf{(d)} Collapse of the data shown in \textbf{b}. Each time trace is scaled using its respective best-fit parameter values. We define $\Delta I^{2\omega}\equiv 1-I^{2\omega}(t)/I^{2\omega}(t<0)$}
    \label{Fig2}
\end{figure*}
From the fits we extract three parameters: (i) the change in SHG intensity immediately following the pump pulse, $I_0$; (ii) the time constant of the recovery, $\tau$; and (iii) the shape parameter, $\beta$. When $\beta<$1, this function is a “stretched” exponential and is asymptotically slower than exponential, whereas when $\beta>$1, it is faster than exponential and referred to as a “compressed” exponential. 
The best-fit values of these parameters are plotted as a function of photoexcitation fluence in Fig.~\ref{Fig2}(c). 

%
As the fluence is increased, $I_0$ first increases approximately linearly before reaching a saturation at $F_{sat}\approx$1.8~mJ/cm$^2$, at which point the low-temperature phase is completely suppressed. 
We note that a significant decrease in $I^{2\omega}$ following photoexcitation is observed only below $T_{MI}$ (see Fig.~\ref{Fig3}(b) and Ref.~\cite{bootstrap}), coinciding with the decrease in $I^{2\omega}$ measured upon warming above $T_{MI}$ (Fig.~\ref{Fig1}(c)).
The photoinduced drop in $I^{2\omega}$ therefore represents a suppression of the low-temperature phase.
Additionally, the relaxation time constant $\tau$ increases from $\sim$100~ps to $\sim$1~ns over the measured fluence range.
Above the maximal 2.4~mJ/cm$^2$ shown here, $\tau$ increases such that the functional form of the recovery cannot be determined within the experimental time window.

Most importantly, we report a strong fluence dependence of the shape parameter $\beta$. 
At low fluences, the time traces are stretched exponentials ($\beta<$1) as is evidenced by the diverging slope as $t\xrightarrow{} 0^+$, followed by a convex relaxation. 
As the fluence is increased, $\beta$ increases and exhibits a crossover near 1.1~mJ/cm$^2$ at which point it surpasses unity and the relaxation changes from stretched- to compressed-exponential form. 
The compressed exponentials ($\beta>$1) exhibit a vanishing slope as $t\xrightarrow{} 0^+$ and an inflection point in the subsequent relaxation.
This crossover indicates a qualitative change in the underlying dynamics that can be tuned by adjusting the photoexcitation fluence. 

It is important to note that time traces of differing fluence reach the same SHG intensity with differing slope. 
The dynamics are therefore not solely determined by the observed macroscopic state.
In the remainder of this work, we identify two physical ingredients that we argue are essential to the microscopic physics underlying this anomalous relaxation.
We then demonstrate that these ingredients can account for our observations by incorporating them into a simple numerical model that qualitatively reproduces the experimental behavior.

The first ingredient that we identify as essential to the PIPT is \textit{spatial inhomogeneity}. 
%
A priori, coexistence of the metallic and insulating regions is anticipated due to the discontinuous nature of the thermal phase transition~\cite{Perez-Salinas2022, McLeod21}.
In the data, the partial suppression of the low-temperature phase at fluences below $F_{sat}$ can be naturally attributed to such phase coexistence. 
And indeed, phase inhomogeneity has previously been reported in this PIPT~\cite{bootstrap}. 
%
For these reasons we take the relaxation to be inhomogeneous in our model. 



The other essential ingredient of this PIPT is \textit{cooperative interactions}, which are implied by the compressed-exponential relaxation.
Unlike their stretched counterparts, compressed exponentials can \textit{not} be well-approximated by sums over simple exponentials. 
%
However, if local relaxation events enhance the relaxation rate elsewhere, faster-than-exponential behavior can be achieved.
It is therefore highly probable that there are cooperative interactions between spatially distinct parts of the system. 
Furthermore, we will see in the model detailed below that the addition of interactions ensures that an important feature of the data is reproduced: namely, that the relaxation is not determined solely by the observed macroscopic state.

We simulate CRO as a 2$D$ array of sites that represent distinct mesoscopic regions of the sample. 
Motivated by the discontinuity of the thermal metal-insulator transition, we restrict the state of each site to be either excited or relaxed, corresponding respectively to the high- and low-temperature phases. 
To emulate the effect of the pump pulse, a random subset of sites is excited at $t=0$, with the fraction of the total sites given by the fluence fraction parameter $f$. Spatial inhomogeneity is therefore built into the model. 
\begin{figure*}
    \includegraphics[scale=1.1]{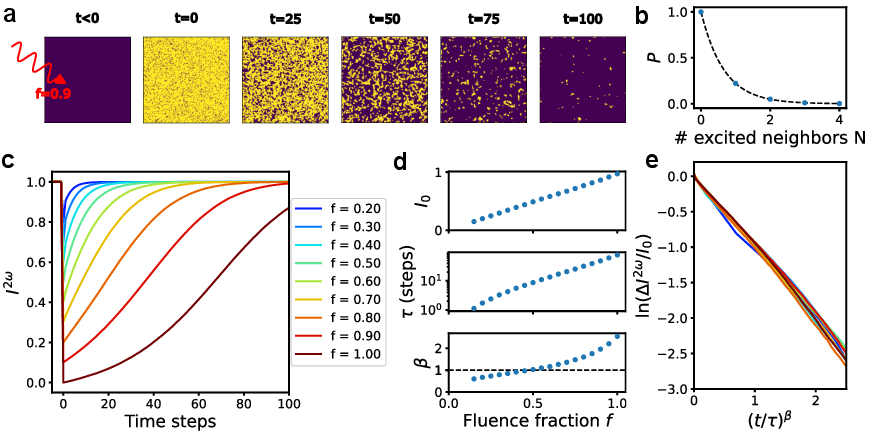}
    \centering
    \caption{Numerical simulation of the PIPT. Results are shown for $s=6$ on a 150x150 array with periodic boundary conditions.\textbf{(a)} The simulated state at select time steps for fluence fraction $f=0.9$. Purple and yellow squares represent the relaxed and excited states respectively. \textbf{(b)} The site relaxation probability per time step as a function of the number of excited nearest neighbors.\textbf{(c)} Simulated time traces for various fluence fractions $f$. $I^{2\omega}$ represents the fraction of sites that are excited at each time point. Each curve is an average over 10 runs of the simulation. \textbf{(d)} The parameters obtained by fitting the simulated time traces in \textbf{c} to Equation~\ref{Eqn1}, plotted versus the fluence fraction. \textbf{(e)} Data collapse of the simulated SHG time traces shown in \textbf{c}, performed in the same manner as in Fig.~\ref{Fig2}(c)}
    \label{Fig4}
\end{figure*}

Cooperative interactions are included in the neighbor-dependent relaxation of the excited states. 
At every time step, each excited site relaxes with probability $P(t+1)=\mathrm{exp}(-sN(t)/4)$, where $N(t)/4$ is the fraction of nearest neighbors that are excited at time $t$.
This expression can be thought to represent an energy barrier to relaxation that increases with the local strain imposed by neighboring sites (see below).
The number $s$ is the only free parameter in the model and controls the strength of the strain-mediated relaxation suppression. 
To correlate the simulations with experimental SHG data, we equate the simulated SHG intensity to the fraction of sites that are in the excited state at each time step.
The resultant simulated SHG time traces are shown in Fig.~\ref{Fig4}(c) for various fluence fractions.


The simulation reproduces the salient qualitative features of the experimental data.
As done for the experimental time traces, each simulated SHG curve is fit to Eq.~\ref{Eqn1}. 
Fig.~\ref{Fig4}(e) demonstrates the strong adherence to the stretched/compressed exponential form, showing the data collapse of the simulated time traces, equivalent to that shown for the experimental data in Fig.~\ref{Fig2}(d).
In addition, the simulation reproduces the fluence dependence of the fit parameters, shown in Fig.~\ref{Fig4}(d).
Most importantly, $\beta$ evolves from less than one to greater than one with increasing fluence, marking a crossover from stretched- to compressed-exponential relaxation.
This simple model therefore demonstrates that a minimal implementation of spatial inhomogeneity and cooperative interactions can account for the observed anomalous relaxation.

The fluence dependence of the relaxation time and of the shape parameter can now be understood within the framework of the model.
As the fluence is increased, the relaxation time $\tau$ increases because there are more inter-site interactions that suppress relaxation. 
%
The origins of the crossover in $\beta$, however, are more subtle. 
Stretched exponentials are well-approximated by sums over simple exponentials with varying time constants~\cite{MAURO18,Ediger1996}.
Therefore, macroscopic probes that average over many microscopic regions may measure stretched exponentials when the microscopic relaxation rate is inhomogeneous~\cite{Richert2002}.
The cooperative interactions, however, favor compressed-exponential relaxation because each site relaxation increases the relaxation probability of its excited neighbors.
This is reminiscent of avalanche-like behaviors that have been associated with compressed-exponential relaxation~\cite{Evenson15,Trachenko21,Song22}.
As the fluence is increased, excited sites become more connected to each other through nearest-neighbor pairs, and the interactions dominate, causing the crossover to $\beta>1$.
In various physical systems like glasses or jamming transitions, compressed exponentials are thought to be related to microscopic interactions, and in particular to the release of internal stresses~\cite{Ruta12,Cipelletti00,Caronna08,Guo09,Wu18,Cipelletti03,Bouchaud01,Ferrero14, Ballesta08,BANDYOPADHYAY06}.
In this experiment, in which the observed relaxation is structural, a strain-mediated interaction mechanism is especially plausible~\cite{Zhao21}. 

\begin{figure}
    \centering
    \includegraphics[scale=0.35]{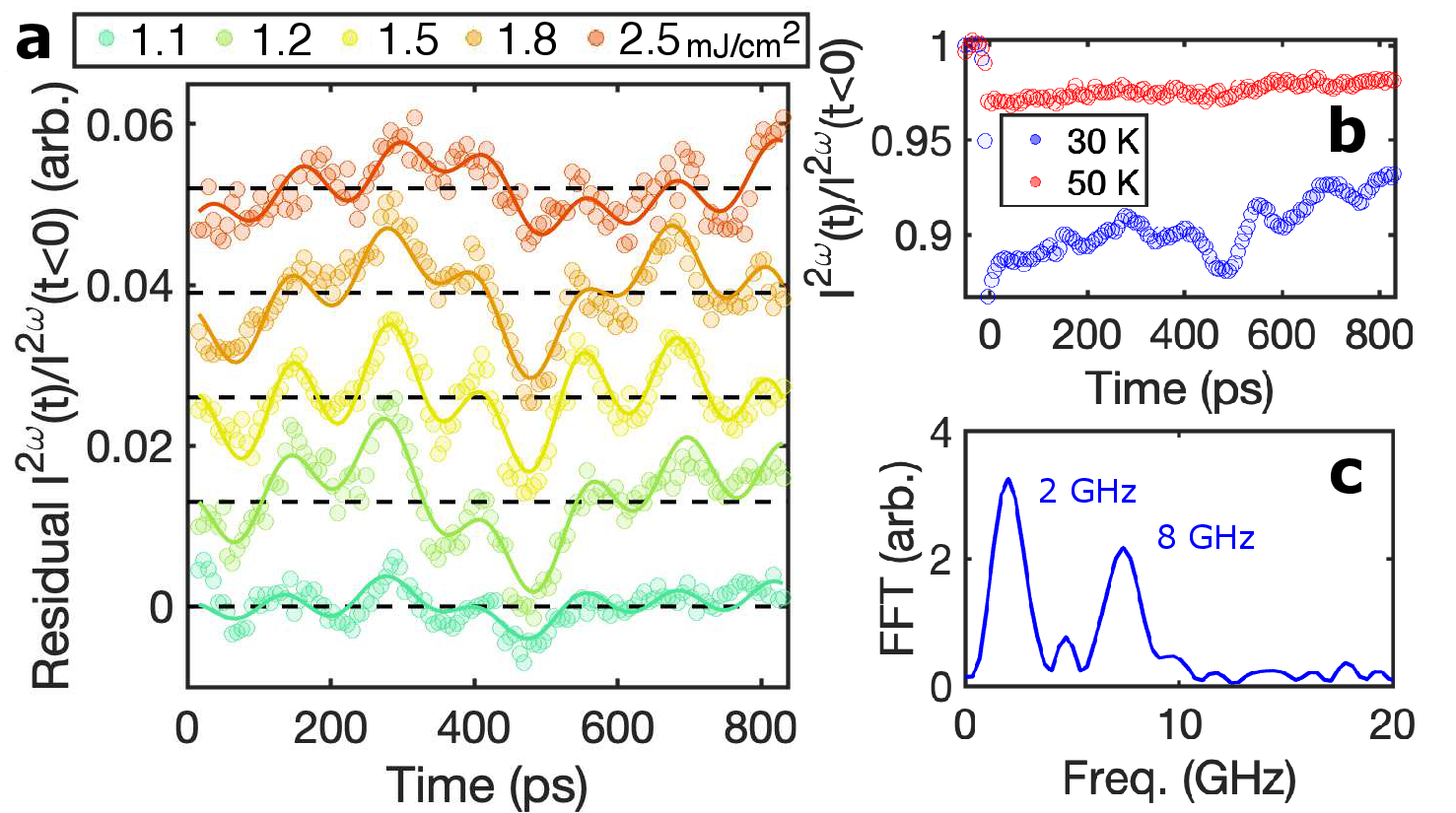}
    \caption{Low-frequency oscillations during relaxation. \textbf{(a)} Residuals from fits shown in Fig.~\ref{Fig2}(a) for select fluences. Solid lines are fits to a sum of three sinusoidal functions with approximate frequencies of 2~GHz, 4~GHz, and 8~GHz. 
    \textbf{(b)} SHG time traces measured above and below $T_{MI}$ = 48~K at T = 30~K and T = 50~K with a 1.3~mJ/cm$^2$ pump pulse.\textbf{(c)} Fast Fourier transform of the residual shown in \textbf{a} for 1.5~mJ/cm$^2$ 
    }
    \label{Fig3}
\end{figure}

Further evidence for a strain-mediated relaxation mechanism comes from the observation of coherent low-frequency oscillations. 
%
In Fig.~\ref{Fig3}(a), we plot the oscillatory component of the relaxation curves shown in Fig.~\ref{Fig2}(b). 
The Fourier transform of these residuals, shown for an incident fluence of 1.5~mJ/cm$^2$ in Fig.~\ref{Fig3}(c), reveals at least two primary frequencies (2 and 8~GHz). 
Given this frequency range and the sensitivity of the SHG probe to the lattice distortions, we attribute these modes to acoustic phonons.
Importantly, these modes appear intimately related to the structural phase transition, as they disappear when the system is heated above $T_{MI}$ (Fig.~\ref{Fig3}(b)).
%
Moreover, the GHz oscillations onset concomitantly with the compressed-exponential relaxation as a function of fluence~\cite{Supp}, suggesting that strain underlies both phenomena.
%
%
Thus, we posit that strain-mediated cooperative interactions both drive the crossover from stretched- to compressed-exponential relaxation and give rise to the observed acoustic modes.

Our work establishes an important connection between the relaxation dynamics of photoexcited solids and that of disordered systems more generally. 
Using ultrafast spectroscopy, we have uncovered a crossover between stretched- and compressed-exponential relaxation in a photoinduced phase transition. 
By supplementing these results with numerical simulations, a comprehensive picture is obtained in which this rich glassy dynamics arises from inhomogeneous relaxation with basic interactions.
We expect that these findings will illuminate the origins of anomalous relaxation in various domains, and will inspire future investigations of glassy dynamics in photoexcited solids.

Research at U.C.L.A. was supported by the U.S. Department of Energy (DOE), Office of Science, Office of Basic Energy Sciences under Award No. DE-SC0023017 (experiment and simulations).
G.C. acknowledges National Science Foundation support via Grant No. DMR 2204811.

\newpage

\title{Supplementary Material to ``Dynamical crossover between stretched- and compressed-exponential relaxation in a photoexcited crystal"}

\author{Tyler Carbin}
\email{tcarbin@g.ucla.edu}
\affiliation{Department of Physics and Astronomy, University of California Los Angeles, Los Angeles, CA 90095, USA}

\author{Joel Herman}
\affiliation{Department of Physics and Astronomy, University of California Los Angeles, Los Angeles, CA 90095, USA}

\author{Xinshu Zhang}
\affiliation{Department of Physics and Astronomy, University of California Los Angeles, Los Angeles, CA 90095, USA}

\author{Yu Zhang}
\affiliation{Department of Physics, University of Colorado at Boulder, Boulder, Colorado 80309, USA}

\author{Hengdi Zhao}
\affiliation{Department of Physics, University of Colorado at Boulder, Boulder, Colorado 80309, USA}

\author{Adrian B. Culver}
\affiliation{Department of Physics and Astronomy, University of California Los Angeles, Los Angeles, CA 90095, USA}
\affiliation{Mani L. Bhaumik Institute for Theoretical Physics, Department of Physics and Astronomy, University of California Los Angeles, Los Angeles, CA 90095}

\author{Rahul Roy}
\affiliation{Department of Physics and Astronomy, University of California Los Angeles, Los Angeles, CA 90095, USA}
\affiliation{Mani L. Bhaumik Institute for Theoretical Physics, Department of Physics and Astronomy, University of California Los Angeles, Los Angeles, CA 90095}

\author{Gang Cao}
\affiliation{Department of Physics, University of Colorado at Boulder, Boulder, Colorado 80309, USA}

\author{Anshul Kogar}
\email{anshulkogar@physics.ucla.edu}
\affiliation{Department of Physics and Astronomy, University of California Los Angeles, Los Angeles, CA 90095, USA}

\date{\today}

\maketitle


\clearpage

\onecolumngrid

\noindent{\Large\textbf{Supplementary Material to ``Dynamical crossover between stretched- and compressed-exponential relaxation in a photoexcited crystal"}}

\tableofcontents

\section{Material Synthesis}

Single crystals were synthesized using an optical floating zone method. Precursors for crystal growth were synthesized using standard solid-state reaction techniques. CaCO$_3$ (Alfa Aesar, 99.99\%) and RuO$_2$ (Colonial Metals, 99.9\%) with molar ratio 1:1 were thoroughly ground and sintered in an MTI KSL1500XS muffle furnace at 900 °C for 24 h. The excess RuO2 was to compensate for volatilization losses during growth. The sintered precursor was then packed into balloons and pressed hydrostatically at 80 MPa into a feed rod ($\sim$7 cm length, $\sim$6 mm diameter) and a seed rod ($\sim$3 cm length, $\sim$6 mm diameter). These rods were sintered again at 1000~$^\circ$C for 24 hours to improve the density.
Crystal growth was carried out in a NEC 2-mirror optical floating-zone furnace equipped with a water-cooled cold trap to capture volatilized RuO2. During growth, a high air flow ($>$0.5 L min$^{-1}$) at 0.3 MPa was maintained to minimize deposition of RuO$_2$ on the shielding glass tube. The feed and seed rods were counter-rotated at 30 rpm to ensure homogeneity of the molten zone. The seed rod was translated at a high speed of 35 mm/h to reduce the RuO$_2$ loss, while the feed-rod speed was actively adjusted to maintain a stable melting zone throughout growth.
The resulting single crystals were characterized by single-crystal X-ray diffraction (SCXRD) and magnetic susceptibility measurements to confirm phase purity, with particular attention to ruling out competing Ruddlesden–Popper phases such as Ca$_2$RuO$_4$.

\section{Rotational anisotropy SHG analysis}

The source term for the radiation emitted from the crystal can be expressed by a multipole expansion: 
\begin{equation}
    \mathbf{S}\propto\frac{\partial^2\mathbf{P}}{\partial t^2}+\nabla\times\frac{\partial\mathbf{M}}{\partial t}+\nabla\frac{\partial^2\mathbf{Q}}{\partial t^2},
\end{equation}
where the first term is the leading-order electric dipole (ED) term followed by the magnetic-dipole (MD) and electric-quadrupole (EQ) terms.
The measured rotational anisotropy (RA) SHG pattern is fit by the leading-order ED contribution,
\begin{equation}
    P_i(2\omega)=\chi_{ijk}E_j(\omega)E_k(\omega),\label{eq:def of chiED}
\end{equation}
where $E_i(\omega)$ is the incident electric field at frequency $\omega$, $P_i(2\omega)$ is the induced polarization at frequency $2\omega$, and a summation over repeated spatial indices is implied.

Enforcing the nonlinear susceptibility $\chi_{ijk}$ to obey the material point group symmetry ($C_{2v}$) as well as the permutation symmetry of indices $j$ and $k$ leaves only five nonzero tensor elements: $\chi_{ccb}=\chi_{cbc}$, $\chi_{aba}=\chi_{aab}$, $\chi_{bcc}$, $\chi_{baa}$, and $\chi_{bbb}$.

The intensity of the second harmonic is proportional to the norm squared of the dot product of $\mathbf{P}(2\omega)$ and the unit vector along the polarization direction of the analyzer $\mathit{\mathbf{e}}(2\omega)$:
\begin{equation}
    I^{2\omega}\propto|\mathit{\mathbf{e}}(2\omega)\cdot \mathit{\mathbf{P}}(2\omega)|^2.
\end{equation}

In the experiment, we rotate the incident light polarization and the analyzer simultaneously, keeping the angle between them fixed. 
The second harmonic intensity takes the following forms for the cases in which the these two polarizations are parallel and perpendicular, respectively:
\begin{equation}
\begin{aligned}
    & I^{2\omega}_{Par}(\theta)=|(2\chi_{aba}+\chi_{baa})\mathrm{cos}(\theta)\mathrm{sin}^2(\theta)+\chi_{bbb}\mathrm{cos}^3(\theta)|^2 \\
    & I^{2\omega}_{Perp}(\theta)=|(2\chi_{aba}-\chi_{bbb})\mathrm{cos}(\theta)\mathrm{sin}^2(\theta)-\chi_{baa}\mathrm{cos}^3(\theta)|^2
\end{aligned}
\label{RAeqn1}
\end{equation}
where $\theta$ is the angle between the incident polarization and the crystallographic $b$ (polar) axis. The solid lines shown in Figure~1b are the results of a simultaneous fit of both the parallel and perpendicular data to these functions.

\section{Additional data}

\begin{figure*}
    \centering
    \includegraphics[scale=0.35]{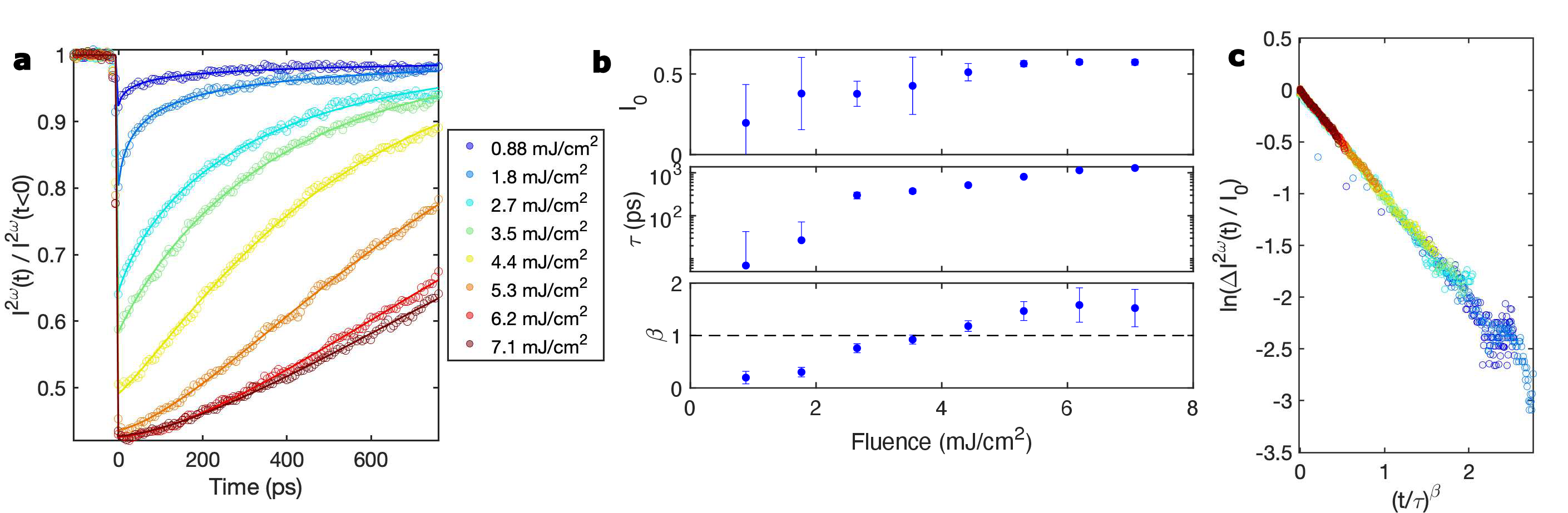}
    \caption{SHG relaxation data equivalent to Fig~2 of the main text, measured at a different location on the sample, with a probe beam of size 35~$\mu$m HWHM and fluence 3.0~mJ/cm$^2$ and a pump beam of size 59~$\mu$m.\textbf{(a)} Time evolution of the normalized SHG intensity $I^{2\omega}$ following photo-excitation for varying pump fluence at a nominal temperature of 4~K (laser heating raises the temperature). Solid lines are fits to Equation~1 of the main text. \textbf{(b)} Best-fit values for each of the fit parameters in Equation~1 as a function of pump fluence. Error bars are 95\% confidence intervals from the fitting procedure.
    \textbf{(c)} Collapse of the data shown in \textbf{(a)}. Each time trace is scaled using its respective best-fit parameter values.}
    \label{ExtraData1}
\end{figure*}

We measured SHG relaxation data as a function of photoexcitation fluence, as shown in Fig.~2 of the main text, at various locations on the sample and with various values of the pump and probe spot sizes. 
A representative dataset is shown in Figure~\ref{ExtraData1}.
Across these measurements, the qualitative features of the data that are essential to our interpretation and that are reproduced by the numerical simulations are maintained. These include the saturation of $I_0$ at high fluence, the rapid increase in $\tau$ with fluence, and the crossover in $\beta$ from less-than to greater-than one.
Features of the data that varied between measurements include the minimum and maximum values of $\beta$ and the amplitude of the GHz oscillations. 

\section{Fitting analysis of time-resolved SHG response}
\subsection{Logistic growth}

Interestingly, we obtain a decent fit of the SHG relaxation data to a model of logistic growth. 
The obtained fits are poorer than those of the stretched/compressed exponential, but they involve only two rather than three independent fit parameters while maintaining decent agreement with the data. 

\begin{figure*}[h]
    \centering
    \includegraphics[scale=0.35]{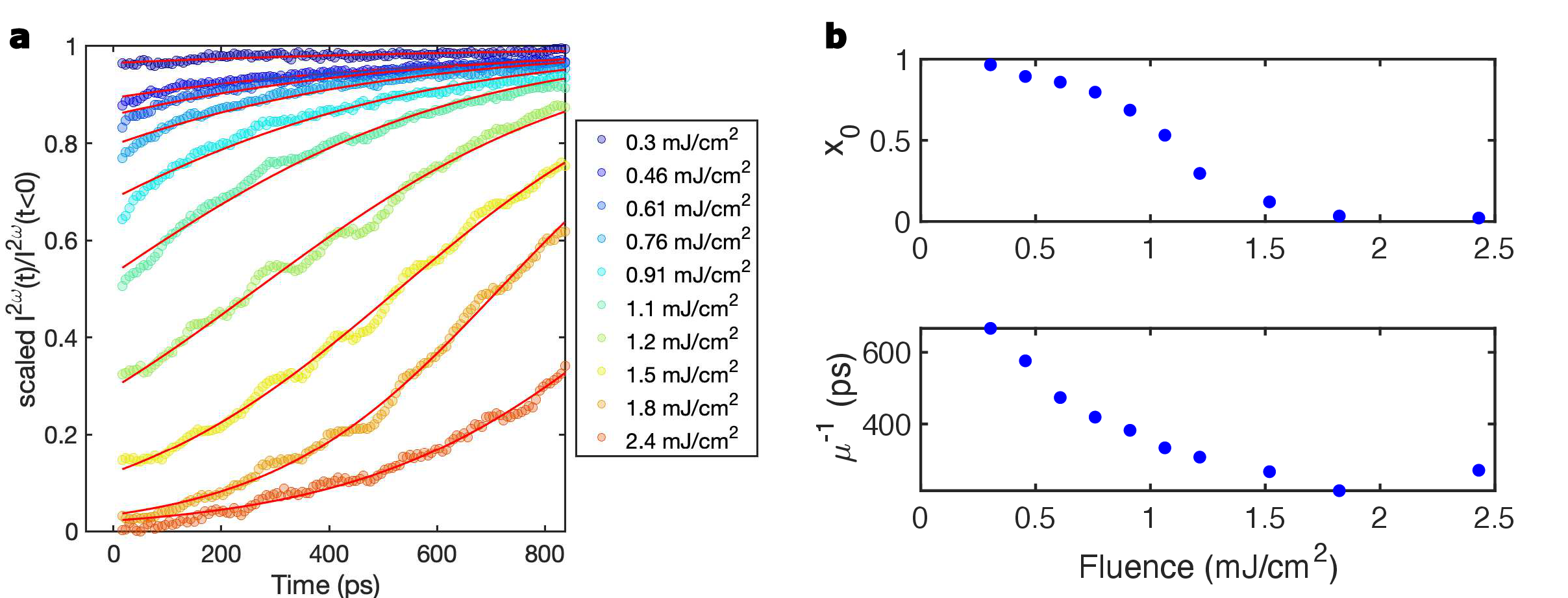}
    \caption{Fits to logistic growth model. \textbf{a} The SHG relaxation data in the main text, rescaled between 0 and 1 on the $y$-axis. The red curves are fits to Equation~\ref{eq:logistic} with $K$=1 fixed. \textbf{b} The corresponding fit parameters as a function of fluence.}
    \label{fig:Logistic}
\end{figure*}

The logistic growth model entails the differential equation
\begin{equation}
    \dot{x}=\mu x \left( 1-\frac{x}{K}\right),
\end{equation}
where $x$ typically represents the population size of a given species in an ecosystem, and $\mu$ and $K$ are the growth rate and the carrying capacity, respectively.
This has the solution:
\begin{equation}
    x(t)=\frac{K}{1+\frac{K-x_0}{x_0}e^{-\mu t}},
    \label{eq:logistic}
\end{equation}
where $x_0$ is the population size at $t$=0.
This model describes growth of a variable $x$ that is moderated by the carrying capacity $K$, but would otherwise be exponential. 
To apply this model to the SHG relaxation data, we first rescale the SHG intensity to traverse between 0 and 1, so that it represents the fraction of the system that is in the insulating state.
As $x$ represents the growing fraction of the system in the insulating phase, it has a strict upper bound of $x$=1. 
We therefore fix the carrying capacity, $K$=1. 
The only remaining fit parameters are then $x_0$ and $\mu$. In Figure~\ref{fig:Logistic}, we present the results of fitting the data in the main text to this logistic model.
Although the fits are poorer than those to the stretched/compressed exponential, the broad success of the fitting is noteworthy given the simplicity of the model and the fact that a single parameter, $x_0$, captures both the starting value of the relaxation curves as well as their concavity.

\subsection{Logarithmic relaxation}
\begin{figure}
    \centering
    \includegraphics[scale=0.4]{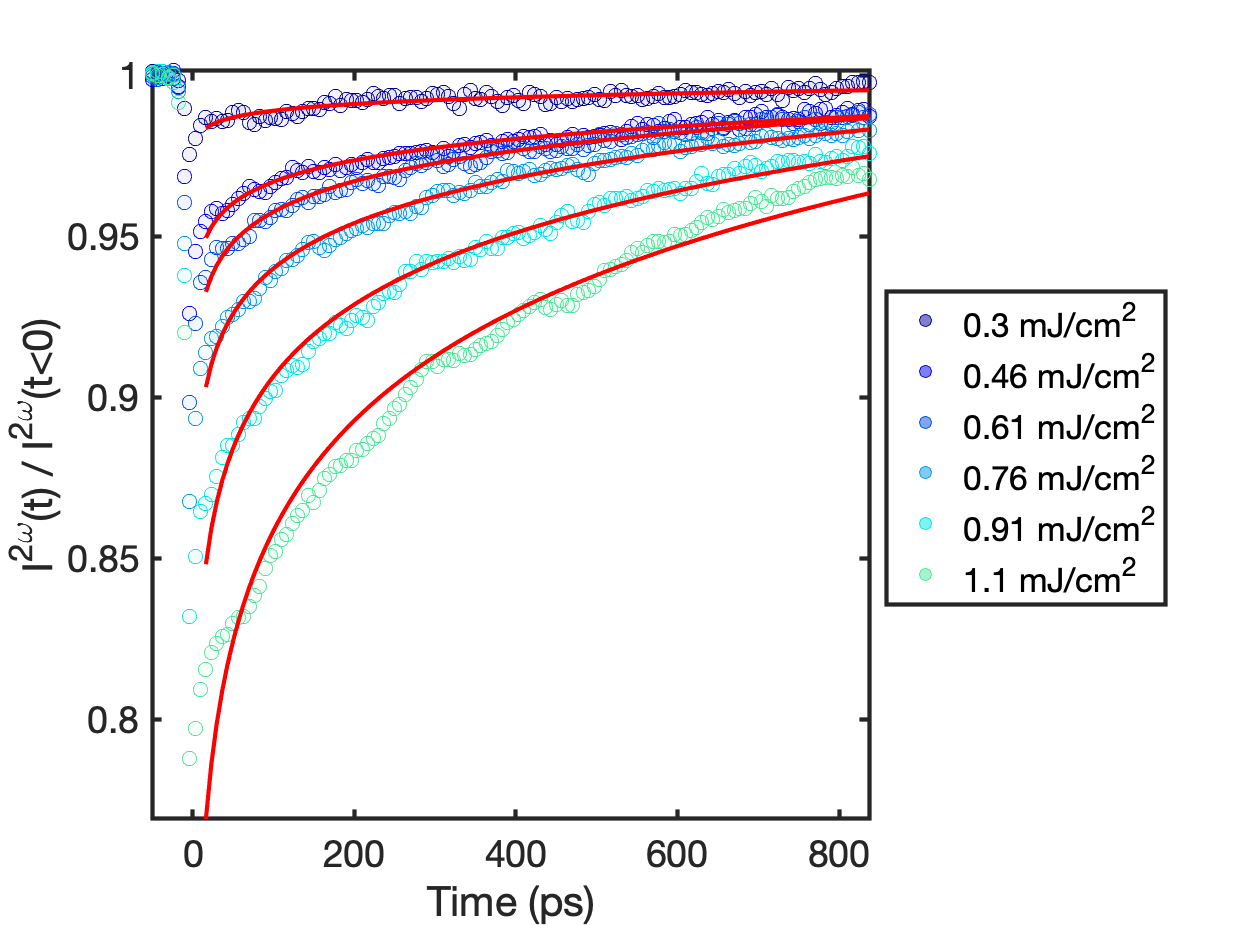}
    \caption{Fits of the data shown in the main text to a logarithmic relaxation given by Equation~\ref{eq:logRec}. Only fluences in the stretched-exponential regime are included.}
    \label{fig:LogFits}
\end{figure}

For the fluences at which the SHG time traces follow a stretched exponential, we present fits to a logarithmic relaxation. 
In particular, we fit the following functional form:
\begin{equation}
    u(t)=I_1+I_0 \mathrm{ln}\left(t/\tau \right).
    \label{eq:logRec}
\end{equation}
The results, shown in Figure~\ref{fig:LogFits} affirm that the stretched exponential is the more appropriate fitting function.

\subsection{Avrami kinetics}

\begin{figure*}[h]
    \centering
    \includegraphics[scale=0.4]{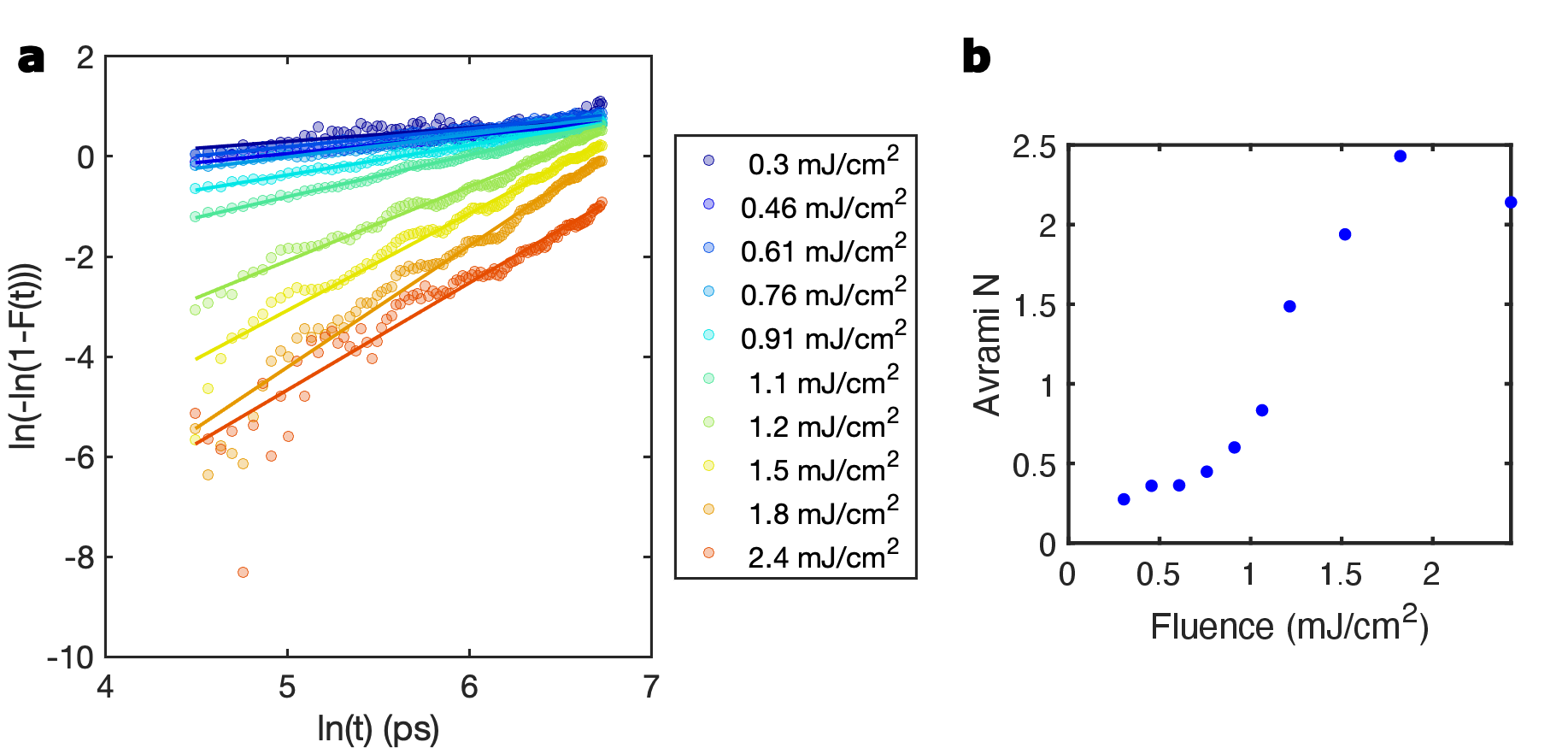}
    \caption{Avrami analysis \textbf{(a)} Rescaling of the data shown in the main text such that the Avrami exponent is given by the slope. Solid lines are linear fits. \textbf{(b)} The extracted slopes from the fits in \textbf{a}, plotted as a function of fluence. }
    \label{fig:Avrami}
\end{figure*}

Many discontinuous phase transitions exhibit kinetics that are governed by nucleation and growth and that are well described by the Avrami equation~\cite{Avrami1,Avrami2,Avrami3}:
\begin{equation}
    f(t)=1-\mathrm{exp}\left(-kt^N\right).
    \label{eq:Avrami}
\end{equation}
Here, $f$ is the fraction of the system that has transformed to the new phase and $t$ is time. 
We rescale the SHG relaxation data in Fig.~\ref{fig:Avrami}(a) so that the Avrami exponent $N$ is given by the slope of each trace. 
The extracted values for $N$ are plotted as a function of fluence in subfigure (b), and agree with the best-fit values of the stretched/compressed exponential shape parameter $\beta$, shown in Fig.~2.

The exponent $N$ can be interpreted within the Avrami model. 
For nucleation that is continuous in time during the relaxation, the Avrami exponent is $N=d/m+1$, where $d$ is the dimension of the system and $m$ is the growth mode parameter that is 1 for ballistic growth and 2 for diffusive growth. In the opposite limit in which the nucleation occurs only at $t\leq0$, the exponent is $N=d/m$. 
A straightforward application of the Avrami model to our experimental data therefore indicates that the dimension and the growth mode change as a function of photoexcitation fluence. 
While such a scenario cannot be ruled out, it is more likely that nucleation and growth as captured by the Avrami equation do not govern the entire dynamics reported here. 
In particular, at fluences below $F_{sat}$, the system is inhomogeneous immediately following photoexcitation, whereas analysis based on the Avrami equation typically assumes a uniform state prior to nucleation.






\subsection{Finite time window}
Due to the length of our delay stage, we can not observe the relaxation to arbitrarily large times $t$ after photoexcitation.
At large photoexcitation fluences at which the relaxation time $\tau$ becomes significant compared to the experimental time window ($\approx$800~ps), the best-fit parameters must be extracted with incomplete information of the relaxation. 
In Fig.~\ref{fig:smallfitrange}, we demonstrate the effects of this time window restriction on the best-fit parameters by fitting the simulated time traces shown in the main text over a shorter time window. 
In subfigure (b), we see that the primary effect of this restricted time range is to decrease the best-fit value for $\beta$ at the largest fluences, reminiscent of that extracted from the experimental data (c.f. Fig.~2(c)).

\section{Temperature effects}
The scenario we have put forward in this work to explain our observations at $T=$4~K is not intended to apply at higher temperatures arbitrarily close to $T_{MI}$. 
In our model, the site relaxation probability is exponentially dependent on the number of excited neighbors. As mentioned in the main text, this may be interpreted as an strain-mediated energetic barrier to relaxation that is linear in the number of excited neighbors.
At the very least, in adapting our model to higher temperature, we may expect to require a temperature-dependent contribution to this relaxation probability, which may be nontrivial, potentially accounting both for the Boltzmann factor and for the temperature dependence of the inter-site coupling. 
More generally, as the temperature is increased, a greater number of relaxation pathways may become available to the system, and the extension of our model to this regime may not be straightforward. 
Nonetheless, our findings may mark a crucial step toward understanding the effects of increased temperature on this dynamics. 

\subsection{Repetition Rate dependence}

\begin{figure*}[h]
    \centering
    \includegraphics[scale=0.35]{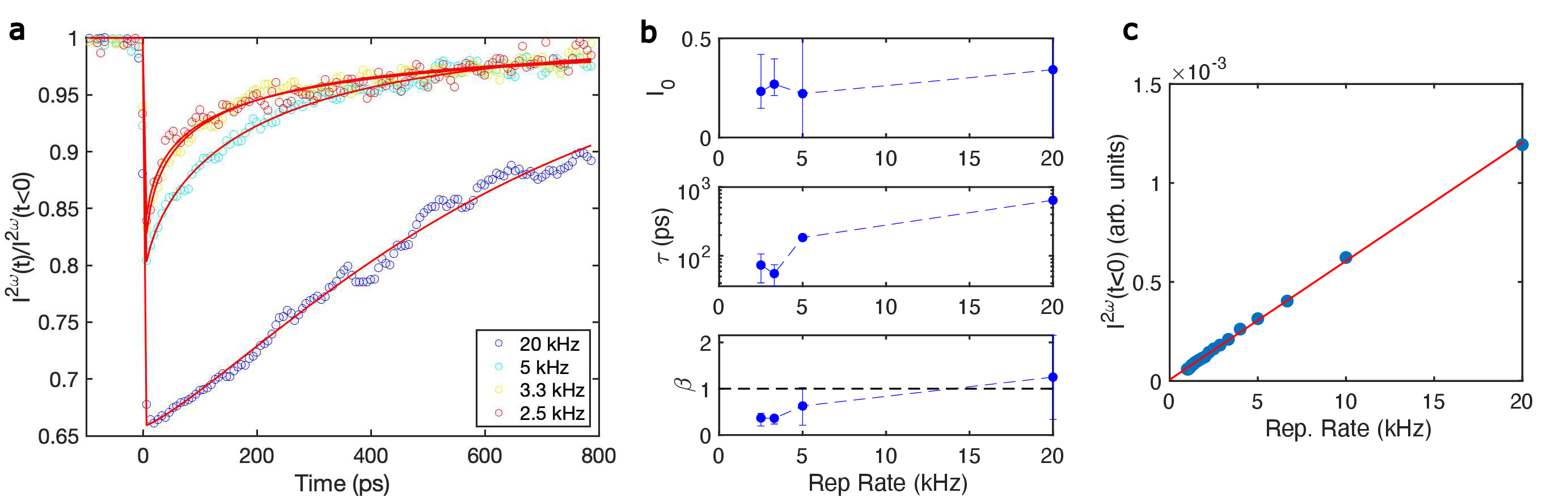}
    \caption{Relaxation of the PIPT as a function of the repetition rate of the pump and probe pulses. 
    \textbf{(a)} The normalized SHG intensity following photoexcitation for various laser repetition rates. The probe beam has a fluence of 1.3~mJ/cm$^2$ and a FWHM of 35 $\mu$m. The pump beam has a fluence of 0.71~mJ/cm$^2$ and a FWHM of 210~$\mu$m. \textbf{(b)} The corresponding fit parameters as a function of laser repetition rate. \textbf{(c)} The SHG intensity measured with the probe arriving slightly before the pump pulse (i.e. $t$ is small and negative) as a function of the laser repetition rate. The pump and probe fluences are 1.3~mJ/cm$^2$ and 2.0~mJ/cm$^2$, respectively, with the same spot sizes as in \textbf{a}. The red line is a linear fit.}
    \label{fig:RepRateTraces}
\end{figure*}

In Figure~\ref{fig:RepRateTraces}, we show that varying the repetition rate of the laser while maintaining fixed fluences for the pump and probe beams can cause notable changes in the relaxation measured by SHG. We argue that the primary effect of changing the repetition rate is the resultant variation in the steady state temperature. 

As the repetition rate is increased, the power imparted on the material is increased, and thus the steady state temperature increases. 
However, increasing the repetition rate can cause nonthermal effects if the dynamics instigated by one pump pulse are not completed by the time the following arrives. 
To ensure that the sample fully recovers between consecutive pump pulses, we fix the probe pulse to arrive just before the pump pulse (i.e. $t$ is small and negative) and vary the repetition rate. 
The resultant SHG intensity as a function of repetition rate is plotted in Figure~\ref{fig:RepRateTraces}c. 
The linear trend of this curve indicates that the sample fully recovers to the insulating state between consecutive excitations by the pump pulse. 
It is therefore most natural to interpret the dependence of the relaxation on the repetition rate as a thermal phenomenon in which the predominant effect of increasing the repetition rate is to raise the steady state temperature. 


\subsection{Transient pump-induced heating}
From the heat capacity, we obtain an upper bound on the transient temperature increase following absorption of the pump pulse. The lattice temperature increases from $T_i$ to $T_f$ according to
\begin{equation}
    q=\int_{T_i}^{T_f}\mathit{dT}c(T),
\end{equation}
where $q$ is the injected energy per unit cell and $c(T)$ is the heat capacity per unit cell, which we determine from Ref.'s~\cite{VARADARAJAN2007402,McCall03}.

The energy per unit cell injected to the sample by the pump pulse is given by
\begin{equation}
\begin{aligned}
    q&=(1-R)F \frac{V_{uc}}{\lambda^{pen}_{1030}} \frac{\int_0^\infty dz e^{-z/\lambda^{pen}_{1030}}(e^{-z/\lambda^{pen}_{1030}})^2}{\int_0^\infty dz e^{-z/\lambda^{pen}_{1030}}} \\
    &= \frac{2}{3}(1-R)F \frac{V_{uc}}{\lambda^{pen}_{1030}},
\end{aligned}
\end{equation}
where $R$~=0.18~\cite{Lee07} is the reflectivity at 1030~nm, $F$ is the fluence, $V_{uc}$ is the volume of the unit cell, $\lambda^{pen}_{1030}$=78~nm is the penetration depth, and the final ratio of integrals accounts for the sensitivity of the SHG probe along the depth direction. 
To determine an upper bound on the final transient temperature $T_f$, we equate these expressions for the injected energy per unit cell. 
The sample is cooled to a nominal temperature of $T_i$= 4~K, but the true initial temperature is strictly greater due to laser heating from the probe beam. 
If the initial temperature $T_i$ is 10~K, the upper bound for the transient temperature following photoexcitation of the maximal fluence, 2.4~mJ/cm$^2$, is 45.6~K, which remains below $T_{MI}$.


\section{Comparison of simulated and experimental results}

\subsection{Fits to simulated time traces}

We present the fits to the simulated SHG time traces shown in the main text. 
\begin{figure}[H]
    \centering
    \includegraphics[width=0.5\linewidth]{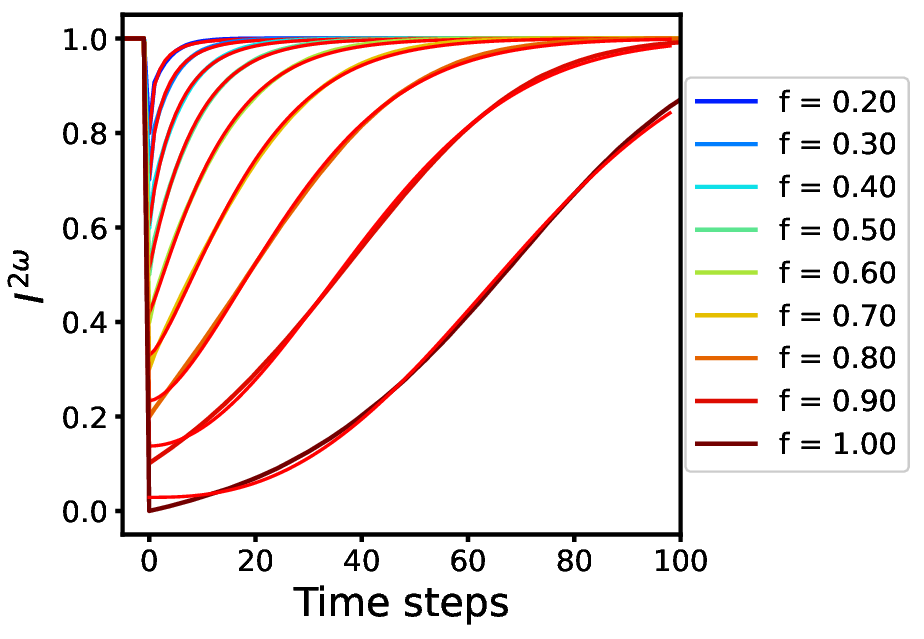}
    \caption{Fits to a stretched/compressed exponential relaxation function, shown in red, overlayed on the simulated time traces, shown in Fig.~3(c). These fits determine the best-fit parameters, shown in Fig.~3(d).}
    \label{fig:enter-label}
\end{figure}

\subsection{Bidirectional dynamics and $I_0$ saturation}

In the numerical simulations presented in the main text, the dynamics are unidirectional: following the pump pulse, excited sites relax but relaxed sites cannot become excited. However, in our previous work~\cite{bootstrap}, we observed the sample to evolve inhomogeneously from the low-temperature phase toward the high-temperature phase (i.e. become excited) immediately following the photoexcitation pulse. 
For simplicity in the simulations presented in the main text, we do not include these dynamics that move the system \textit{away} from equilibrium because they occur on a much shorter timescale than the subsequent relaxation that is the focus of this work ($<$2~ps compared to hundreds of ps).

However, an addition of a simple type of dynamics that moves the system away from equilibrium may shed light on an apparent difference between the simulations and the experiment. 
The experimental data exhibit a saturation of $I_0$ with increasing fluence that is absent from the simulated time traces in the main text. 
This lack of a saturation in the simulations is a direct consequence of the design of the model; saturation is only achieved at fluence fraction $f$ = 1, i.e. when all sites are excited by the pump pulse. In the experiment, one can apply more light than is required to excite the entire sample, but in the simulation only 100\% of the sites can be excited.

One way to observe a saturation in the simulations is to include dynamical rules that move the system away from equilibrium in addition to those that enable the relaxation.
As already mentioned, both types of dynamics are observed in the PIPT. In Fig.~\ref{fig:simsbothdirs}, we present the results of simulations with the same dynamics as those in the main text but with an additional rule that counteracts the relaxation: namely, if a relaxed site has greater than a given number of excited nearest neighbors, it becomes excited. We see that for fluences slightly less than one, these dynamics can increase $I_0$ in a way that resembles the saturation of $I_0$ observed in the experiment. 




\subsection{Depth effects}

The SHG probe has an effective penetration depth that is dependent on the penetration depth at both the fundamental and second harmonic frequencies~\cite{bootstrap}:
\begin{equation}
    \lambda^{\mathrm{pen}}_\mathrm{eff}=\frac{\lambda^{\mathrm{pen}}_{515 }\lambda^{\mathrm{pen}}_{1030}}{\lambda^{\mathrm{pen}}_{515}+\lambda^{\mathrm{pen}}_{1030}}=30~\mathrm{nm}.
\end{equation}

\begin{figure}
    \centering
    \includegraphics[scale=0.5]{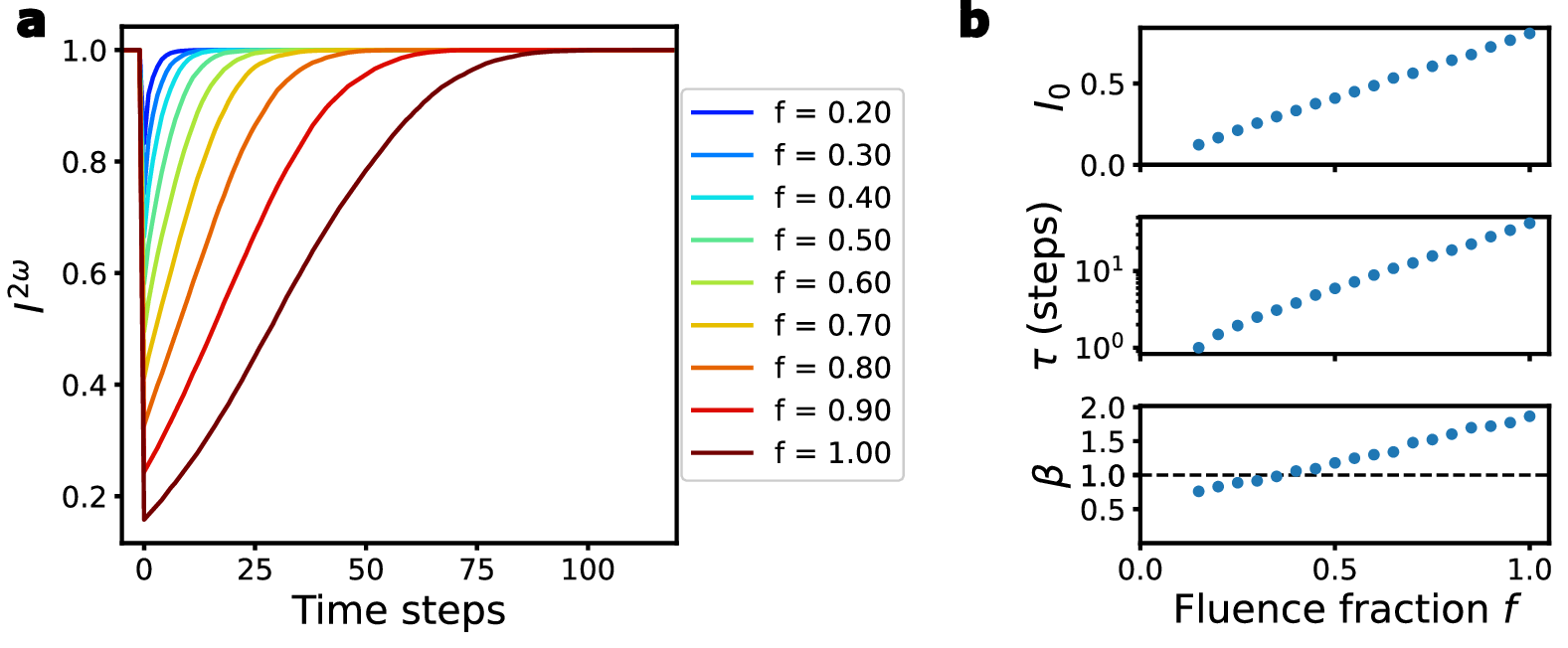}
    \caption{\textbf{(a)} Simulated time traces on a 30x30x30 grid in which the finite penetration depths of the pump and probe are accounted for. 
    The ratio of the penetration depths of the pump and probe in the simulations is given by the experimental ratio $\lambda_{\mathrm{eff}}^{\mathrm{pen}}/\lambda^{\mathrm{pen}}_{1030}=$2.4.
    The penetration depth of the pump is set to 80 sites. \textbf{(b)} The stretched/compressed-exponential best-fit parameters.}
    \label{fig:pendepthsims}
\end{figure}

\begin{figure}[h]
    \centering
    \includegraphics[scale=0.5]{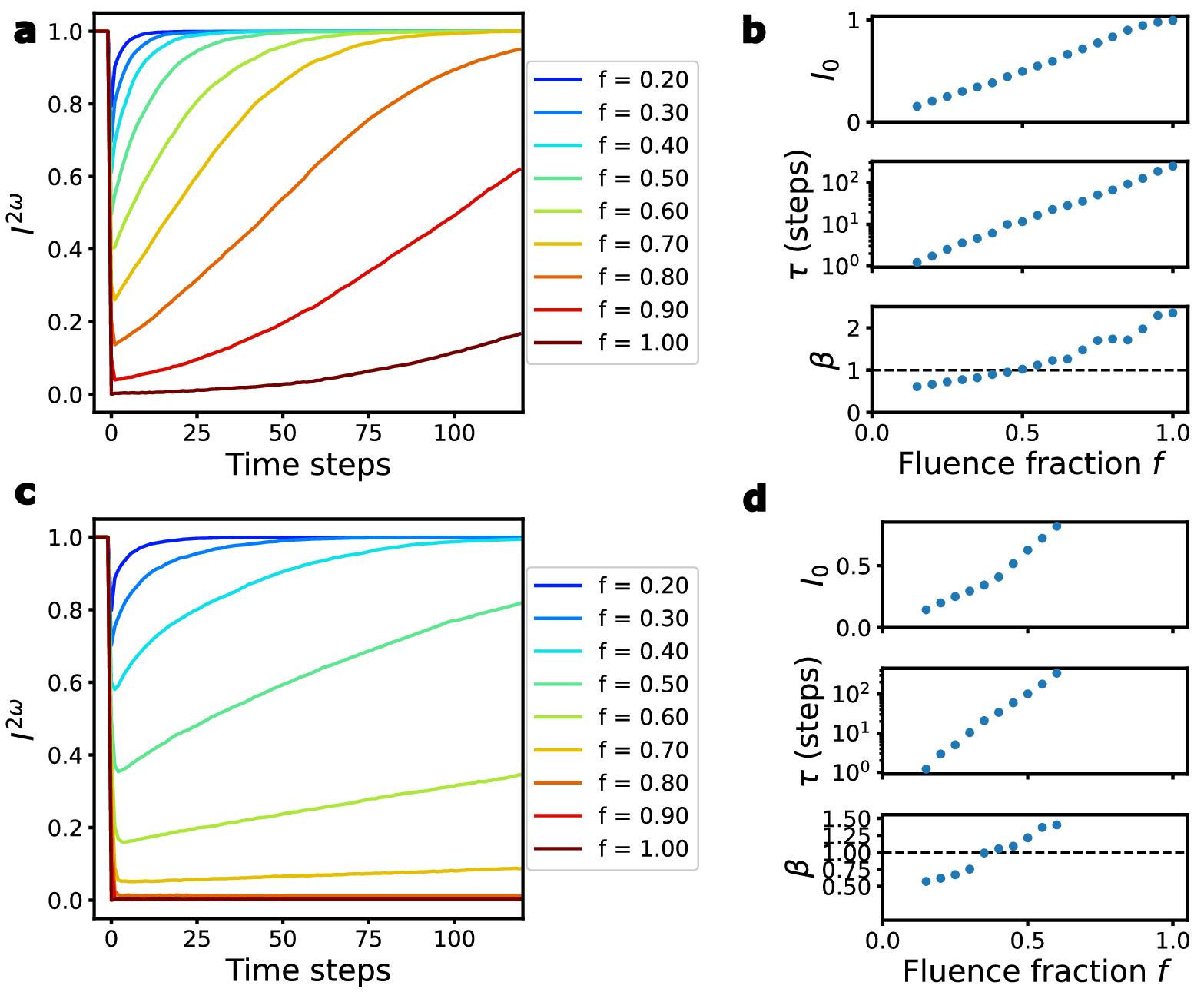}
    \caption{Results of simulations that are identical to those in the main text but with the additional dynamical rule that 0 (relaxed) sites become 1 (excited) with probability 1 if $n_c$ or more of their nearest neighbors are excited. \textbf{(a,b)} $n_c=4$ and \textbf{(c,d)} $n_c=3$. In both cases, results are shown for $s$=6 on a 150x150 grid with periodic boundary conditions.}
    \label{fig:simsbothdirs}
\end{figure}

\begin{figure}[h]
    \centering
    \includegraphics[scale=0.5]{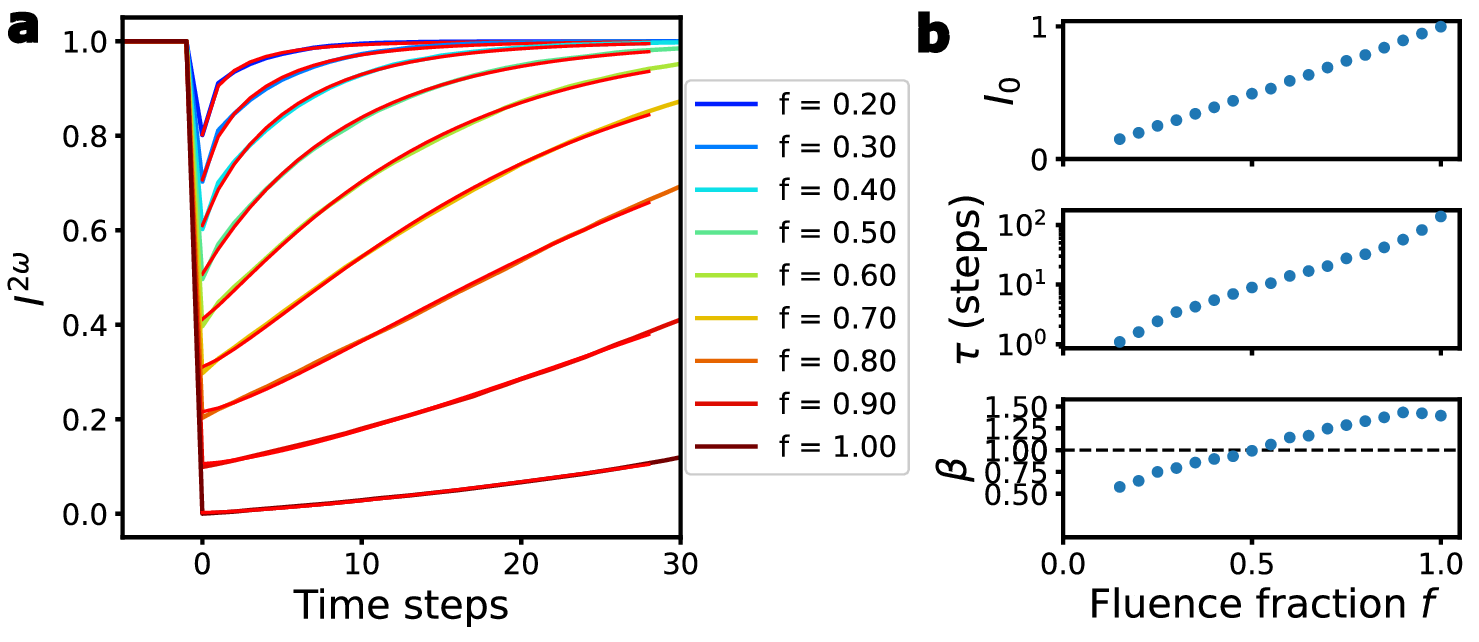}
    \caption{\textbf{(a)} Fits to the simulated time traces shown in the main text in which the time range of the fits has been decreased. \textbf{(b)} Corresponding fit parameters as a function of fluence.}
    \label{fig:smallfitrange}
\end{figure}

This is significantly less than the penetration depth of the pump pulse, $\lambda^{\mathrm{pen}}_{1030\mathrm{nm}}=$~78~nm. For this reason, we have chosen to present the results of a simulation with two spatial dimensions in the main text. For completeness, in Fig.~\ref{fig:pendepthsims} we present the results of a simulation with three spatial dimensions in which the finite penetration depths of the pump and probe beams are accounted for. 
To simulate the penetration depth of the probe, the contribution of each site to the SHG signal is exponentially suppressed along the depth direction. Similarly, the photoexcitation is suppressed exponentially along the depth direction. The ratio of the penetration depth of the pump to that of the probe is determined by the experiment: $\lambda_{\mathrm{eff}}^{\mathrm{pen}}/\lambda^{\mathrm{pen}}_{1030}=$2.4.
The qualitative behavior of the simulated time traces is largely unaffected. 

\section{GHz Oscillations}

\begin{figure*}[h]
    \includegraphics[scale=0.5]{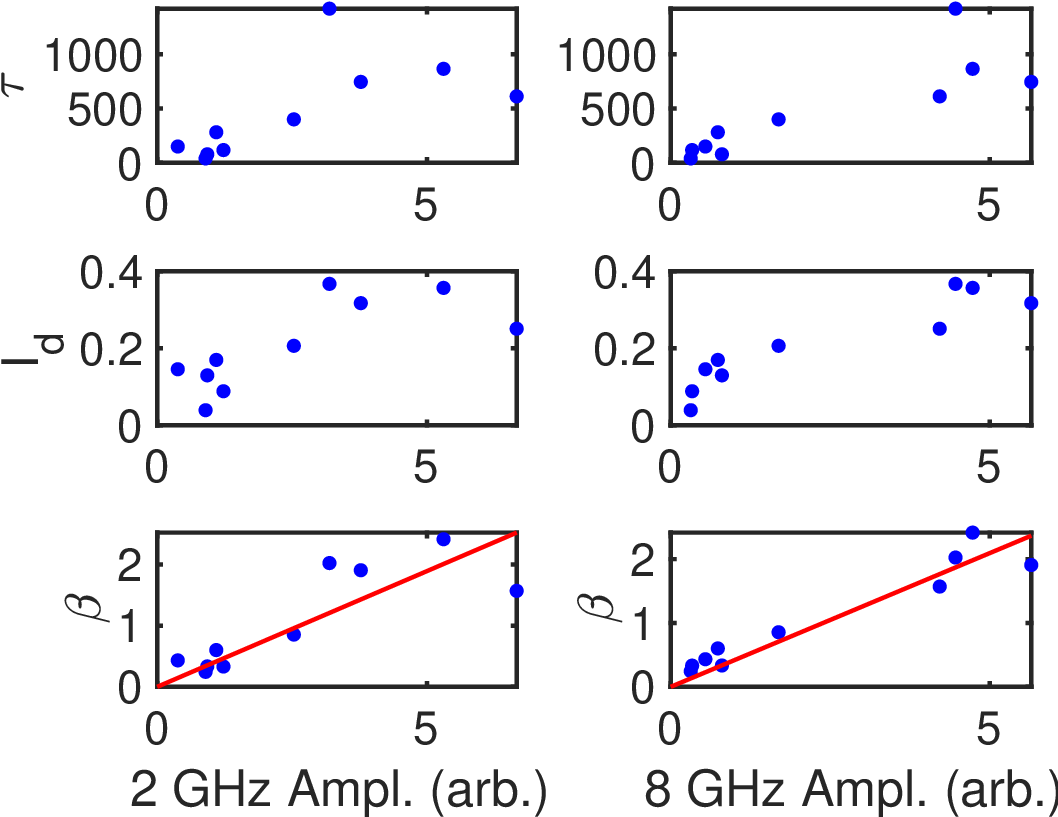}
    \centering
    \caption{Correlations between amplitudes of 2~GHz and 8~GHz oscillations and stretched/compressed exponential fit parameters for the data presented in the main text. Phonon amplitudes are calculated by integrating the FFT of the residuals shown in Figure~4 of the main text over a 1~GHz range centered at 2 and 8~GHz. Red lines are linear fits with zero intercept.}
    \label{fig:PhononParamsCorrs}
\end{figure*}

In Figure~\ref{fig:PhononParamsCorrs}, we show the correlations between the 2 and 8~GHz oscillation amplitudes and the fit parameters of the stretched/compressed-exponential relaxation. 
The correlation between the oscillation amplitudes and the shape parameter $\beta$ supports the claim that the cooperative interactions underlying the compressed-exponential relaxation are intimately related to the coherent oscillations. 

\bibliography{references}

\end{document}